\documentclass[print,epsfig,twocolumn,floats,showpacs]{revtex4}
\usepackage{graphicx}
\usepackage{dcolumn}
\usepackage{bm}
\usepackage{amssymb}
\usepackage{amsmath}
\usepackage{epsfig}
\usepackage{CJK}

\begin{document}

\title{Superfluidity and Stabilities of a Bose-Einstein condensate with periodically modulated interatomic interaction}
\author{Shao-Liang Zhang}
\affiliation{Key Laboratory of Quantum Information,
University of Science and Technology of China, Hefei, Anhui
230026, China}
\author{Zheng-Wei Zhou}
\affiliation{Key Laboratory of Quantum Information,
University of Science and Technology of China, Hefei, Anhui
230026, China}
\author{Biao Wu}
\email{wubiao@pku.edu.cn}
\affiliation{International Center for Quantum
Materials, Peking University, 100871 Beijing, China}

\begin{abstract}
We study theoretically the superfluidity and stability of a  Bose-Einstein
condensate (BEC) whose interatomic scattering length is periodically modulated
with optical Feshbach resonance.  Our numerical study finds that the properties
of this periodic BEC are strongly influenced by the modulation strength.
When the modulation strength is small, only the Bloch waves close to the Brillouin zone
edge suffer both Landau and dynamical instabilities.  When the modulation strength is
strong enough, all Bloch waves become dynamically unstable. In other words,
the periodic BEC loses its superfluidity completely.
\end{abstract}

\maketitle

\section{Introduction}
Even though superfluidity is one of the most important macroscopic
quantum phenomena,  it could only be found in liquid helium before
1995. Since the realization of Bose-Einstein condensation in atomic
gases in 1995, we have now in experiment another superfluid, the
Bose-Einstein condensate (BEC)~\cite{rBEC}.  This new superfluid
shares many interesting properties with superfluid helium,  such as
critical velocity~\cite{rSF} and quantized vortices~\cite{rVot}.  At
the same time, there are also some interesting properties unique to
this gaseous superfluid; for example, there is no roton excitation
in BECs.  In particular, since a gas is easily compressible, one can
modulate its density with an optical lattice to create a periodic
superfluid. The properties of this periodic superfluid have been
studied extensively both theoretically~\cite{r001} and
experimentally~\cite{exp001}. Dynamical instability, which is absent
in a homogenous superfluid, was discovered and found to play a
dominant role in destroying superfluidity in a periodic
superfluid~\cite{r001,exp001}.

Now another type of periodic superfluid can be created: with optical
Feshbach resonance (OFR)~\cite{rFR},  the interatomic interaction
(or scattering length) of a BEC can be modulated periodically  in
space with laser beams~\cite{r006}. This was already demonstrated in
experiment~\cite{r000}. This BEC with periodically modulated
interaction (PMI) is different from the usual BEC in an optical lattice,
where the atoms feel an external periodic potential (PP).
The most important difference between these two periodic BEC systems
is that the BEC with PMI has no linear periodic
counterpart. As a result, the widely used single-band approximation
for a periodic system appears not applicable for a BEC with PMI.
Furthermore,  while the BEC in PP has a Mott-insulator phase~\cite{rOpT,rOpL},
the BEC with PMI should not have the Mott phase.   In this work we
study  the superfluidity and other related physical properties in a BEC with PMI.

We focus on the  case where the laser beam for OFR is applied only
along one direction. In this case,  the BEC can be
described by the following Gross-Pitaevskii equation
\begin{equation}
i\hbar\frac{\partial\Psi}{\partial t}=
-\frac{\hbar^2}{2m}\nabla^2\Psi+[V_1+V_2\cos(2k_Z x)]|\Psi|^2\Psi\,,
\end{equation}
where $V_1$ and $V_2$ are positive parameters that can be tuned
experimentally by changing laser power and
detuning~\cite{rFR,r006,r000}; $m$ is the atomic mass and $k_Z$ is
the wave number of the laser beam.   There have already been some
theoretical attempts to find soliton solutions in such a
system~\cite{r007,rNOL}. In this work we study the superfluidity of
this periodic system. This is equivalent to examining the stability of
a flow described by a Bloch wave~\cite{WuShi}. The Bloch wave
solutions of this BEC system for the lowest Bloch band are found
numerically. Their Landau and dynamical stabilities are examined by
computing the Bogoliubov excitations. We find that when the periodic
modulation strength $V_2$ is small,  only Bloch waves close to the
Brillouin zone edge are unstable. When the modulation strength
becomes large enough, all Bloch waves in the lowest band become
dynamically unstable. This means that the periodic modulation of the
scattering length can cause a BEC to lose its superfluidity
completely. When this happens,  the periodic Bose system is neither a
superfluid nor a Mott insulator; the BEC may collapse into many
solitons as suggested by an early study~\cite{r005}. In contrast,
for a BEC in PP,  the Bloch state near the Brillouin zone center is
always stable no matter how strong the periodic modulation
is~\cite{r001}.

The paper is organized as follows. In
Sec.\uppercase\expandafter{\romannumeral2}, we present the basic
theoretical framework within which the BEC system is treated.
In Sec.\uppercase\expandafter{\romannumeral3}, the lowest
two Bloch bands for this periodic BEC system are presented
and their physical meaning is discussed. In
Sec.\uppercase\expandafter{\romannumeral4}, we study the Landau
instability and dynamical instability of the Bloch waves and discuss
the superfluidity of this system.
In Sec. \uppercase\expandafter{\romannumeral5},  the results
in the previous section are discussed in the context of the experiment.
The paper is summarized in Sec.\uppercase\expandafter{\romannumeral6}.

\section{Theoretical Framework}
We consider the one-dimensional case, where the scattering length is
modulated only in one direction by optical Feshbach resonance
and the lateral motion of the BEC can be either ignored or confined.
In this case the Gross-Pitaevskii equation becomes one-dimensional
\begin{equation}
i\frac{\partial}{\partial t}\psi=
-\frac{1}{2}\frac{\partial^2}{\partial x^2}\psi+(c_1+c_2\cos
x)|\psi|^2\psi\,. \label{eq1d}
\end{equation}
Here the energy unit is $4\hbar^2k^2_Z/m$, and the length unit is
$1/2k_Z$. The wave function $\psi$ is in units of $\sqrt{n_0}$, where
$n_0$ is the averaged BEC density;
$c_1=\frac{mn_0V_1}{4\hbar^2k^2_Z}$, and
$c_2=\frac{mn_0V_2}{4\hbar^2k^2_Z}$.  For convenience, we call $c_1$
the uniform strength and $c_2$ the modulation strength.

We are interested in the superfluidity and stability of a flow in this
system. For a homogeneous BEC, the flow is described by a plane wave.
For a periodic system, the flow is represented by a Bloch wave. The Bloch wave has the form
$\psi(x)=e^{ikx}\varphi_k(x)$, where $\varphi_k(x)$ is  of period $2\pi$,
it satisfies the time-independent Gross-Pitaevskii equation
\begin{equation}
\mu\psi= -\frac{1}{2}\frac{d^2}{d x^2}\psi+(c_1+c_2\cos
x)|\psi|^2\psi\,. \label{bloch1d}
\end{equation}
We use the numerical method proposed in Ref.\cite{wu_njp} to find
these Bloch waves.

Once a Bloch wave solution is found, its stability
is examined.
We add a small perturbation to the Bloch wave $\varphi_k(x)$,
\begin{equation}
\delta\varphi_{k,q}(x)=u_k(x,q)e^{iqx}+v^{\ast}_k(x,q)e^{-iqx}
\end{equation}
where $q$ is in the range $[-1/2, 1/2]$ and represents the mode
of perturbation. The energy deviation caused by the perturbation is
\begin{equation}
\delta
E_k=\int^\infty_{-\infty}dx(u^\ast_k,v^\ast_k)M_k(q)\left(\begin{array}{c}u_k \\
v_k
\end{array}\right),
\end{equation}
where
\begin{equation}
M_k(q)=\left(\begin{array}{cc}
\mathcal{L}(k+q) &(c_1+c_2\cos x)\varphi^2_k\\
(c_1+c_2\cos x){\varphi^\ast_k}^2 & \mathcal{L}(-k+q)
\end{array}\right)
\end{equation}
with
\begin{equation}
\mathcal{L}(k)=-\frac{1}{2}\left(\frac{\partial}{\partial
x}+ik\right)^2-\mu+2(c_1+c_2\cos x)|\varphi_k|^2\,.
\end{equation}
If the matrix $M_k(q)$ has negative eigenvalues, it means that
there are some perturbations $\delta\varphi_{k,q}(x)$  that can lower
the system energy.  This energetic instability is related to the superfluidity
of the system and we call it the Landau instability as
this is the essence behind Landau's theory of superfluidity. The Landau instability is sometimes
called a thermodynamical instability.

We also consider the dynamical evolution of the system after the perturbation
$\delta\varphi_{k,q}$. For such a small perturbation, the dynamical
Eq.(\ref{eq1d}) can be linearized and becomes
\begin{equation}
i\frac{\partial}{\partial t}\left(\begin{array}{c}u_k \\
v_k\end{array}\right)=\sigma M_k(q)\left(\begin{array}{c}u_k \\
v_k\end{array}\right), ~~~ \sigma=\left(\begin{array}{cc}I
& 0
\\ 0 & -I\end{array}\right)
\end{equation}
If  matrix $\sigma M_k(q)$ has complex eigenvalues, the
system becomes unstable and will damp during dynamical evolution.
We call this instability dynamical instability.

Landau instability and dynamical instability can be discussed
in a more coherent way within the framework of Bogoliubov excitations.
The eigenvalues of matrix $\sigma M_k(q)$
can be divided into two groups,  phonon modes and antiphonon
modes; only phonon modes are physical and are usually called Bogoliubov
excitations~\cite{wu_njp}.
It can be proved that matrix $M_k(q)$ having a negative eigenvalue
is equivalent to $\sigma M_k(q)$ having phonon modes of
negative energy~\cite{ChenZhu}.  Therefore, the system has
Landau instability when some of the phonon modes have negative
energies and it has dynamical instability when some of the phonon modes
are complex.  As the phonon modes are related to the superfluidity of a Bose
system~\cite{LandauBook}, it is clear from this perspective that
Landau instability and dynamical instability are clearly related to each other
and are just two ways of destroying superfluidity.  As we show,
both instabilities are present in this periodic BEC.

\section{Nonlinear Bloch bands}
\begin{figure}
\includegraphics[width=1.0\columnwidth]{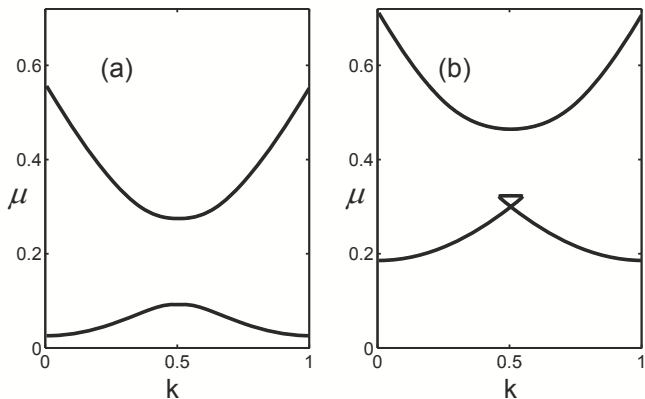}
\caption {The lowest two Bloch bands of a BEC with its interatomic
interaction periodically modulated, $c_2=0.1$: (a) $c_1=0.05$, (b)
$c_1=0.2$. $\mu$ is in units of $4\hbar^2k^2_Z/m$ and
$k$ is in units of $2k_Z$, where $k_Z$ is the wave vector
the laser beam.} \label{fig:LZ}
\end{figure}

Because of the periodic modulation of interatomic interaction, this BEC has
Bloch wave solutions.  These Bloch waves  can in general be found by numerically
solving Eq.(\ref{bloch1d}). However, near the edge of the Brillouin
zone($k\approx 1/2$),  we can use a two-mode
approximation and assume the Bloch state is of the form
$\varphi_{\tilde{k}}(x)\approx
ae^{i(\tilde{k}-\frac{1}{2})x}+be^{i(\tilde{k}+\frac{1}{2})x}$
($|a|^2+|b|^2=1$, $|\tilde{k}|\ll 1$ and $\tilde{k}=k-1/2$). Plugging
the trial wave function into Eq.(\ref{bloch1d}),  we can get a quartic
equation\cite{r004}
\begin{equation}
w^4+2gw^3+(g^2-h^2-1)w^2-2gw-g^2=0\,,
\label{quartic}
\end{equation}
where $g=c_1/c_2$, $h=\tilde{k}/c_2$,
$w=1/(2ab)$,  and the chemical potential is
$\mu=1.5c_1+0.5c_2(w+1/w)$.  It can be shown that
near $\tilde{k}=0$,  Eq.(\ref{quartic}) has only two real solutions
when $g\leq 1$ ($c_1\leq c_2$); it has four real solutions
when $g> 1$ ($c_1> c_2$). This means that
the Bloch band of this nonlinear periodic system has a loop structure
at the edge of the Brillouin zone when $g>1$. This is confirmed by
our numerical computation, as shown in Fig.\ref{fig:LZ}.  In Fig.\ref{fig:LZ} (b),
where $c_1>c_2$, the chemical potential $\mu$ has a clear loop at the edge of
the Brillouin zone.

The loop structure in the Bloch band in Fig.\ref{fig:LZ} (b) is a
manifestation of superfluidity in the system. We can consider the
homogeneous case $c_2=0$. In this case, the BEC is a superfluid with
critical velocity $v_c=\sqrt{c_1}$.  Now we slowly turn on the
periodic modulation by increasing $c_2$ to a small value. This small
periodic modulation can be regarded as a perturbation when $c_1$ is
large. The Bragg scattering caused by this periodic perturbation
should not destroy a superflow moving with velocity $v=1/2$, which
is the velocity at the Brillouin zone edge when $c_2\sim 0$. The
nonzero slope of the Bloch band at the edge is an indication of
this robustness of superfluidity. When the periodic modulation
becomes very strong as in Fig.\ref{fig:LZ}(a), the Bragg scattering
can eventually disrupt the superflow as indicated by the zero slope
at the zone edge in Fig.\ref{fig:LZ}(a). A similar loop structure has
been found in many different systems~\cite{loop}. Note that the loop structure
of the energy band has its interesting many-body counterpart, a net
of narrow avoided crossings in energy levels~\cite{avoidcross}.

\begin{figure}[t]
\includegraphics[width=0.98\columnwidth]{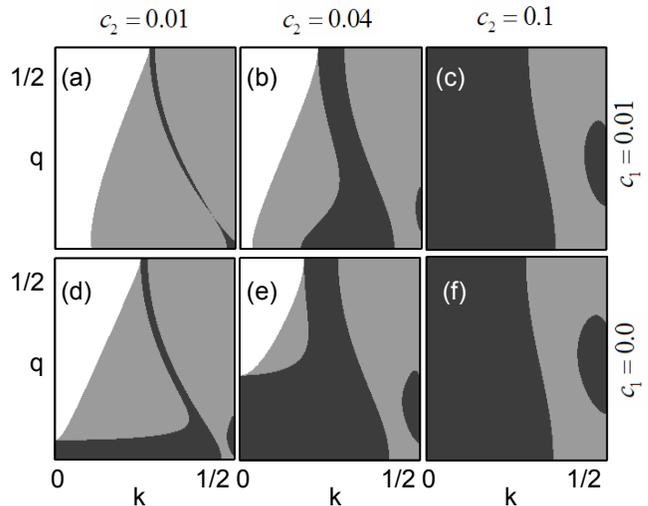}
\caption {Stability phase diagrams of BEC Bloch states for different values of
the uniform strength $c_1$ and  modulation strength $c_2$.
$k$ is the wave number of the Bloch state and $q$ is the
wave number of the perturbation mode.  In the white area, the Bloch state
is a local  energy minimum; in the shaded(light or dark) area,   the Bloch state
has negative excitation energy and Landau instability;  in
the dark shaded area, the Bloch state is dynamically unstable.
$k$ and $q$ are in units of $2k_Z$, where $k_Z$ is the wave vector of
the laser beam.}
\label{fig:AP}
\end{figure}
\section{Superfluidity and instability}
For all the Bloch waves found in the lowest band, we have examined
their superfluidity and stability by numerically computing their
Bogoliubov excitations with $\sigma M_k(q)$.   The results are shown
in the stability phase diagram of Fig.\ref{fig:AP}. For a point
$(k,~q)$ in the figure, there are three possibilities: (i) If
it is in the white region, it means that the Bloch wave $\varphi_k$
is a local energy minimum relative to the perturbation mode $q$.
(ii) If it falls into the black area, the Bloch wave
$\varphi_k$ is dynamically unstable relative to the perturbation
mode $q$. Any perturbation containing mode $q$ will cause the system
to evolve dynamically away from state $\varphi_k$ with an
exponential growth. (iii) If the point $(k,~q)$ lies in the
gray region, the Bloch wave $\varphi_k$ is not a local energy
minimum but dynamically stable relative to the perturbation mode
$q$.  Note that we have only plotted the results for $k>0$ in Fig.
\ref{fig:AP} as the system is symmetric with respect to time
reversal and the results are  the same for $k$ and $-k$.

The stability phase diagram in Fig.\ref{fig:AP}(a) is very similar
to the one for a BEC in PP~\cite{r001}. However, as
$c_2$ increases or $c_1$ decreases,  the phase diagram begins to
have new features. The most prominent is that the black area
(dynamical instability) spreads into the region with $k<1/4$ and
eventually reaches $k=0$. For a BEC in PP, the black
area is restricted in the region with $k>1/4$. It is also clear from
the figure that the boundaries  of the gray area and the black area
coincide once the black area reaches $k=0$. This means that once the
Bloch state at $k=0$ becomes unstable, it has both Landau
instability and dynamical instability. In this case, the ground
state of this system is no longer a Bloch wave. In contrast,
the ground state of a BEC in PP is always a Bloch wave.

\begin{figure}[b]
\includegraphics[width=0.8\columnwidth]{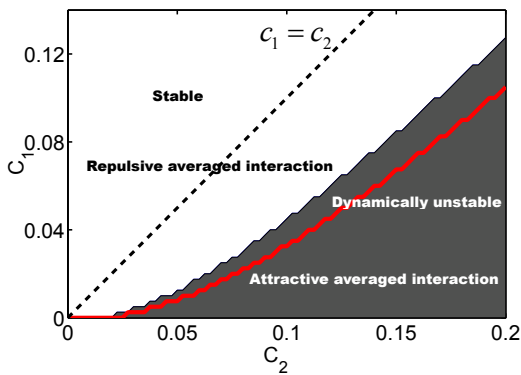}
\caption{(Color online)The stability phases of the Bloch state at $k=0$ in the
space of the interaction parameters $c_1$ and $c_2$. The Bloch wave
state is stable in the white area; it is dynamically unstable in the
black areas. On the solid (red) line,  the  averaged interatomic
interaction is zero; below it, the averaged interaction becomes
negative; and above it, the averaged interaction is positive.}
\label{fig:PHC1C2}
\end{figure}

There is a qualitative way to understand why the Bloch state at
$k=0$ becomes dynamically unstable. When $c_1<c_2$, the interaction
between atoms becomes attractive for some parts of the system. As
$c_2$ increases, a larger portion of the system becomes attractively
interacting.  Eventually, the overall interaction of the system,
indicated by the averaged interaction over one period
$V=\int^{2\pi}_0(c_1+c_2\cos{x})|\psi|^4 dx$, becomes negative. We
expect that this is the underlying reason that the Bloch state at
$k=0$ becomes dynamically unstable. This is confirmed by our
numerically computation. In Fig.\ref{fig:PHC1C2}, we have marked out
the stability regions for the Bloch state at $k=0$ in the parameter
space of $c_1$ and $c_2$. We find numerically that the Bloch state
at $k=0$ is dynamically unstable in the black area in
Fig.\ref{fig:PHC1C2}. The solid(red) line is the dividing line between averaged
positive interaction and averaged negative interaction.
From Fig.\ref{fig:PHC1C2} we can see the solid(red) line is lower
than the stability boundary for the Bloch  state.
It means that the Bloch state is always unstable when the averaged
interaction is negative. However, this is not the only reason that
the Bloch state becomes unstable as there is an area where
the averaged interaction is positive while the Bloch state is unstable.
This is yet to be fully understood.

We now approximate the periodic  BEC with overall negative interaction with
a homogeneous BEC with attractive interaction strength $c$. For a homogeneous
BEC, the flow is described by plane waves $e^{ikx}$. In this case, the matrix
$\sigma M_k(q)$ is a $2\times 2$ matrix. The phonon excitations are easily
obtained. For the state with $k=0$,  they are
\begin{equation}
\epsilon_{\rm ph}=\sqrt{q^2c+q^4/4}\,.
\end{equation}
Since $c$ is negative, the phonon excitation is imaginary for small
$q$. This is in fact the feature seen in
Figs.\ref{fig:AP}(c)-\ref{fig:AP}(f): the Bloch state at $k=0$ is
dynamically unstable against the perturbation of mode $q=0$. This
approximation result further confirms that the averaged
interaction plays a dominant role in this system of a BEC  with PMI
and is the key factor in stability and superfluidity of the BEC system.

\section{experimental perspective}
\begin{figure}
\includegraphics[width=1.0\columnwidth]{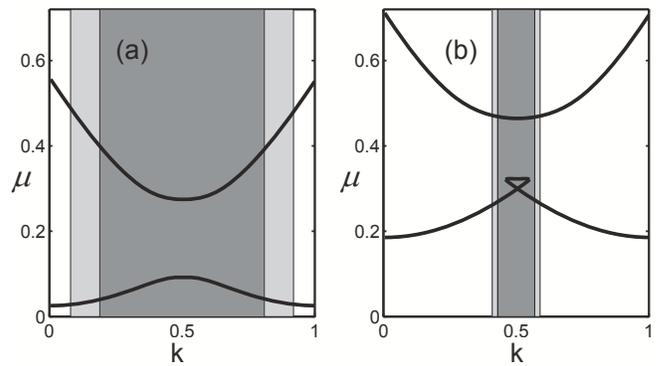}
\caption {Stabilities of the Bloch waves in the lowest Bloch bands
shown in Fig. 1. The Bloch states in the black area are dynamically
unstable, and the Bloch states in the shaded area have Landau
instability, $c_2=0.1$: (a) $c_1=0.05$, (b) $c_1=0.2$.
$\mu$ is in units of $4\hbar^2k^2_Z/m$ and
$k$ is in unit of $2k_Z$, where $k_Z$ is the wave vector of
the laser beam.}
\label{fig:LS}
\end{figure}

The superfluidity and dynamical instability has been explored
experimentally for a BEC in PP~\cite{exp001}. A similar experimental
scheme can be used to study superfluidity and dynamical instability
of this periodic BEC system. There exists  no fundamental
technical barrier.

In a real experiment, there is no way to control the perturbation
mode. The controlled perturbation and the uncontrollable noises in
the experiment should contain all the possible modes of $q$.  Therefore,
for an experimentalist, a Bloch state $\varphi_k$ is unstable if it
is unstable against any of the perturbation modes. For the Bloch
waves in the lowest band in Fig.\ref{fig:LZ}, we have marked out
their stabilities with gray and black shadings  in Fig.\ref{fig:LS}.
It is clear from the figure that the Bloch states near the Brillouin
zone center are more stable and more Bloch states become stable as
the uniform strength $c_1$ increases.

\begin{figure}[!t]
\includegraphics[width=1.0\columnwidth]{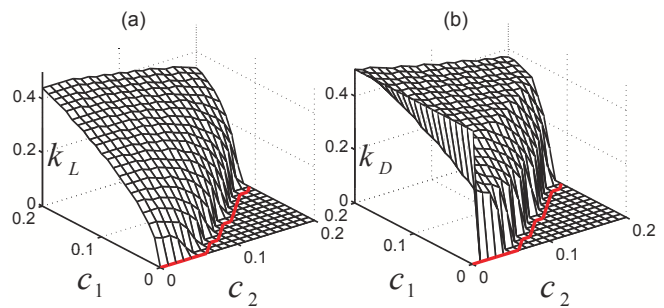}
\caption {(Color online)The critical values of $k_L$ and $k_D$ as a function of
$c_1$ and $c_2$. $k_L$ is the critical Bloch wave number for the Landau
instability and $k_D$ is the critical Bloch wave number for the
dynamical instability. The (red) lines are where
$k_L$ and $k_D$ turn to zero. The two (red) lines in (a) and (b) are identical. $k_L$
and $k_D$ are in units of $2k_Z$, where $k_Z$ is the wave vector of
the laser beam.}
\label{fig:KLD}
\end{figure}

There exist two critical values of $k$ in Fig.\ref{fig:LS}, $k_L$
and $k_D$. A BEC Bloch wave $\varphi_k$ with $k$ in the range
$[k_L, 1/2]$ has Landau instability; it has dynamical instability if
its $k$ is in the range $[k_D, 1/2]$. Both critical values $k_L$ and
$k_D$ vary with $c_1$ and $c_2$, and this dependence is shown in
Fig.\ref{fig:KLD}. As seen  in Fig.\ref{fig:KLD}(a), $k_L$ increases
with the increase of $c_1$ and the decrease of $c_2$. In
Fig.\ref{fig:KLD}(b),  $k_D$ is discontinuous at the edge of
$c_2=0$. This is due to that at $c_2=0$, the system loses the
periodicity and it has no dynamical instability. On the two (red)
lines Fig.\ref{fig:KLD}, both $k_L$ and $k_D$ turn zero. The two (red)
lines are the same. This reflects a fact that we already mentioned:
when the Bloch wave at $k=0$ has Landau instability, it is also
dynamically unstable.

\section{summary}

We have studied a BEC with its scattering length
periodically modulated in one direction. In this
periodic BEC system, the flows are
represented by Bloch waves and the energy has band structure. When
the uniform repulsive interaction is larger than the periodic
modulation strength, the band structure has a loop at the edge of
the first Brillouin zone. We have also studied the stabilities of
these Bloch waves. When the modulation strength is weak, this system
is similar to a BEC in PP. When the modulation strength is
strong enough, even the  Bloch state at the Brillouin zone center becomes
unstable. This means that the BEC loses it superfluidity completely.

\section*{ACKNOWLEDGEMENTS}
This work was funded by the National Natural
Science Foundation of China (Grants No. 11174270 and No. 60921091),
the National Basic Research Program of China (Grant No. 2011CB921204),
the China Postdoctoral Science Foundation No. 2011M501384, the
Fundamental Research Funds for the Central Universities (Grant No.
WK2470000006), and the Research Fund for the Doctoral Program of Higher
Education of China (Grant No. 20103402110024). B. W. is supported by the NBRP
 of China (Grants No. 2012CB921300 and No. 2013CB921900) and the NSF of China (
 Grants No. 10825417, No. 11274024, and No.
 11128407), and the RFDP of China (Grants No. 20110001110091). Z. -W. Z.
gratefully acknowledges the support of the K. C. Wong Education
Foundation, Hong Kong.

\end{document}